\documentclass[11pt,a4paper]{article}
\pdfoutput = 1
\usepackage{jcappub}
\usepackage{bm}
\usepackage{bbm}
\usepackage{amsmath}

\renewcommand\({\left(}
\renewcommand\){\right)}
\renewcommand\[{\left[}
\renewcommand\]{\right]}

\newcommand{\GF}{G_{\rm F}}
\newcommand{\bp}{{\bf p}}
\newcommand{\bx}{{\bf x}}
\newcommand{\br}{{\bf r}}
\newcommand{\bv}{{\bf v}}
\newcommand{\bk}{{\bf k}}
\newcommand{\bK}{{\bf K}}

\newcommand{\sH}{{\sf H}}
\newcommand{\sLambda}{{\sf\Lambda}}
\newcommand{\sM}{{\sf M}}
\newcommand{\sF}{{\sf F}}

\newcommand{\wc}{\omega^{\rm c}}
\newcommand{\ws}{\omega^{\rm s}}

\begin{document}
\subheader{\hfill MPP-2018-224, TIFR-TH-18-25}

\title{Normal-mode Analysis for Collective Neutrino Oscillations}

\author[a,b]{Sagar~Airen,}
\author[b]{Francesco~Capozzi,}
\author[b,c]{Sovan~Chakraborty,}
\author[d]{Basudeb~Dasgupta,}
\author[b]{Georg~Raffelt,}
\author[b]{and Tobias~Stirner}

\affiliation[a]{Indian Institute of Technology Bombay,
Powai, Mumbai 400076, India}

\affiliation[b]{Max-Planck-Institut f\"ur Physik (Werner-Heisenberg-Institut),\\
  F\"ohringer Ring 6, 80805 M\"unchen, Germany}

\affiliation[c]{Department of Physics, Indian Institute of Technology
  Guwahati, Guwahati 781039, India}

\affiliation[d]{Tata Institute of Fundamental Research,
  Homi Bhabha Road, Mumbai 400005, India}

\emailAdd{sagar.airen@iitb.ac.in}
\emailAdd{capozzi@mpp.mpg.de}
\emailAdd{sovan@iitg.ac.in}
\emailAdd{bdasgupta@theory.tifr.res.in.com}
\emailAdd{raffelt@mpp.mpg.de}
\emailAdd{stirner@mpp.mpg.de}

\abstract{In an interacting neutrino gas, collective modes of flavor
coherence emerge that can be propagating or unstable.
We derive the general dispersion relation
in the linear regime that
depends on the neutrino energy and angle distribution.
The essential scales are the vacuum oscillation
frequency $\omega=\Delta m^2/(2E)$, the neutrino-neutrino
interaction energy $\mu=\sqrt{2}\GF n_\nu$, and the matter
potential $\lambda=\sqrt{2}\GF n_e$. Collective modes
require non-vanishing~$\mu$ and may be
dynamical even for $\omega=0$ (``fast modes''), or
they may require~$\omega\not=0$ (``slow modes''). The growth
rate of unstable fast modes can be fast itself (independent
of $\omega$) or can be slow (suppressed
by $\sqrt{|\omega/\mu|}$). We clarify the role of
flavor mixing, which is ignored for the identification of
collective modes, but necessary to trigger collective flavor
motion. A~large matter effect is needed to provide
an approximate fixed point of flavor evolution,
while spatial or temporal variations
of matter and/or neutrinos are required as a trigger, i.e., to
translate the disturbance provided by the mass term to
seed stable or unstable flavor waves.
We work out explicit examples to illustrate these points.
}

\maketitle

\pagebreak

\section{Introduction}
\label{sec:introduction}

The early universe, collapsing stellar cores, or neutron-star mergers
provide environments where neutrinos are so dense that they can have a
strong refractive impact on the dynamics of flavor evolution. In
contrast to ordinary matter, neutrinos tend to develop coherence
between different flavors, producing a flavor-off-diagonal refractive
effect which in turn can cause large flavor conversion
\cite{Pantaleone:1992eq,Duan:2010bg}. The interplay
between flavor coherence and flavor conversion can become
self-accelerating, leading to self-induced flavor conversion. For
appropriate neutrino distributions, such run-away solutions can exist
even in the absence of neutrino mixing (``fast flavor conversion''),
assuming the instability is triggered by a suitable seed.

On the level of a kinetic treatment, the neutrino distribution as a
function of momentum $\bp$ and flavor is described by the usual
occupation numbers, generalized to a matrix structure $\varrho_{\bp}$
in flavor space \cite{Dolgov:1980cq, Barbieri:1990vx}. In this
mean-field description, coherence between different $\bp$ modes is
assumed not to build up, whereas flavor coherence for a given momentum
is followed explicitly in the form of the off-diagonal elements of
$\varrho_\bp$. The diagonal elements are the usual occupation
numbers $f_{\nu_\ell,\bp}$ for the flavors $\nu_\ell=\nu_e$,
$\nu_\mu$ and $\nu_\tau$. In general, the matrix $\varrho_{\bp}$ can
be larger, to include additional degrees of freedom such as spin and
charge parity to encode spin, spin-flavor, or neutrino-antineutrino
correlations \cite{Giunti:2014ixa, Lim:1987tk, Volpe:2013jgr,
  Serreau:2014cfa, Kartavtsev:2015eva, Dobrynina:2016rwy,Vlasenko:2013fja},
but here we stick to the simplest case of flavor
correlations alone.

In the limit of ultra-relativistic neutrinos, the space-time evolution
of the $\varrho_\bp$ matrices is governed by a Boltzmann kinetic
equation of the form \cite{Vlasenko:2013fja, Rudzsky:1990, Sigl:1992fn, Sirera:1998ia,
  Yamada:2000za, Cardall:2007zw, Stirner:2018ojk}
\begin{equation}\label{eq:EOM1}
  (\partial_t+\hat\bp\cdot\partial_\bx)\varrho_\bp=
  -i[\sH_\bp,\varrho_\bp]+{\cal C}[\varrho]\,,
\end{equation}
where $\hat\bp$ is a unit vector in the direction of $\bp$. Moreover,
$\partial_\bx$ is understood as a gradient vector with regard to the
spatial variables. In the Liouville operator on the left-hand side
(lhs) we have ignored a term causing a drift of momenta by external
forces such as cosmic or gravitational redshift or
deflection. On the right-hand side
(rhs), the collision term includes source and sink terms by
charged-current or pair processes as well as collisions between
neutrinos and particles of the ambient medium or among neutrinos. The
collision term includes all momentum modes, so on this level the
equation of motion (EOM) is generically nonlinear.  Coherent flavor
evolution is engendered by the commutator term on the rhs where the
Hamiltonian matrix $\sH_\bp$ depends on the matrix $\sM^2$ of squared
neutrino masses, on the density and distribution of background
particles, and on the $\varrho_\bp$ matrices themselves. It is these $\varrho_{\textbf{p}}$ inside the Hamiltonian, that arise
on the refractive level, that are the source of nonlinearity with which we are concerned in studies of collective neutrino oscillations.

Equation~\eqref{eq:EOM1} is surprisingly hard to solve in situations
of practical astrophysical interest. Even without the flavor
oscillation term, neutrino transport in numerical core-collapse or
neutron-star merger calculations is the most expensive part and
requires various approximations. Flavor oscillations themselves
involve very fast time scales, so a combined brute-force numerical
solution is out of the question. Even post-processing flavor evolution
on a given astrophysical background model defies numerical
solutions. One problem is the emergence of unstable collective modes
that can break the symmetries of the original system. Moreover,
depending on the numerical scheme, spurious unstable modes can
appear and can dominate the numerical result \cite{Sarikas:2012ad,
  Morinaga:2018aug}. Moreover, even in a stationary background
model we are not assured of the stationarity of the solutions for
$\varrho_\bp$ \cite{Abbar:2015fwa, Dasgupta:2015iia,Capozzi:2016oyk}.

If there were no nonlinearity, the flavor dynamics would be independent for each $\bp$ and could be found by solving the corresponding one-dimensional problem. In this case, flavor evolution, signified by the accumulation of a relative phase between the mass eigenstate components of a neutrino mode $\varrho_{\textbf{p}}$, occurs exactly proportional to its displacement along its trajectory in space. It is due to the nonlinearity that flavor and kinematic degrees of freedom evolve differently and the dispersion relation for the flavor waves, obtained via normal-mode analysis, is different from the kinematic one.

To develop a deeper understanding of possible space-time dependent
solutions of equation~\eqref{eq:EOM1}, and in particular to identify
run-away modes, one approach is to study the linearized version of
this equation and look for its normal modes. Depending on wavevector
and frequency, these modes can simply propagate like waves or show
spatial or temporal instabilities. While a normal-mode analysis of
such an equation would seem like the first thing one should do, it is
only recently that one has begun to look at collective flavor
oscillations from this perspective \cite{Izaguirre:2016gsx, Capozzi:2017gqd}.

For a long time one pictured collective flavor conversion as a
stationary phenomenon. Neutrinos
were taken to be emitted from a source, typically the neutrino sphere
of a supernova core, and one asked for the evolution as a function of
distance \cite{Duan:2006an, EstebanPretel:2007ec}.
However, going beyond this simple ``bulb model'' of neutrino
emission with its identical angle distributions for all flavors reveals
the existence of ``fast flavor modes,'' i.e., collective behavior
on a scale defined by the neutrino-neutrino interaction
energy $\mu\sim\sqrt{2}\GF n_\nu$ rather than the vacuum oscillation
frequency $\omega=\Delta m^2/(2E)$
\cite{Sawyer:2005jk, Sawyer:2008zs, Sawyer:2015dsa, Chakraborty:2016lct, Dasgupta:2016dbv, Dasgupta:2017oko}.
Moreover, quite generically one
needs to worry about situations where neutrinos stream in all directions,
so flavor evolution is not trivially a problem of evolution along
some pre-defined spatial direction. Therefore, one is motivated to consider
an interacting neutrino gas, more or less homogeneous on the scales
of perhaps a few meters, and consider possible forms of flavor evolution
irrespective of a simple boundary condition --- on such scales the picture
of a ``neutrino sphere'' makes no physical sense. Therefore, one is naturally
led to worry about general space-time dependent
collective solutions of the EOM and the question
of how they would be triggered.

So far this dispersion-relation approach was applied in the two-flavor
context to the relatively simpler case of fast flavor conversion, where
neutrino masses and mixing are irrelevant except for providing seeds
for fast flavor instabilities.  Here we extend this approach to
include ``slow oscillations,'' driven by neutrino masses, as well as to a
three-flavor framework. We will see that all previously studied
stability-analysis examples are special cases of such a more general
mode analysis. In a simple one-dimensional model of the interacting
neutrino gas (``colliding beams'') we will work out explicit examples
for the interplay between fast and slow modes. Moreover, we will address
the question of seeding coherent flavor motion by the mass term, which is
the only source of flavor violation and thus the only possible source
of collective flavor waves. We clarify the role of non-uniformity of the
matter and/or neutrino density to couple general flavor waves to the
mass term, which by itself is perfectly symmetric
and thus has no overlap with the inhomogeneous
flavor modes.

\pagebreak

\section{Equation of Motion}
\label{sec:eom}

As a first step we set up the different elements entering the
EOM~\eqref{eq:EOM1} and note once more that we work in the limit of
ultra-relativistic neutrinos. In this case masses enter only in the
vacuum oscillation term, whereas otherwise we can take neutrinos to
move with the speed of light~\cite{Stirner:2018ojk}. We denote the
neutrino velocity vector as $\bv=\hat\bp=\bp/|\bp|$ and define a velocity
four-vector as $v^\mu=(1,\bv)$ for every $\bp$ mode. Ignoring
henceforth the collision term, we may thus write the
EOM for flavor oscillations in the form
\begin{equation}\label{eq:EOM2a}
  v^\alpha\partial_\alpha\varrho_\bp=-i[\sH_\bp,\varrho_\bp]\,,
\end{equation}
where a summation over $\alpha=0,\ldots,3$ is implied.
The Hamiltonian matrix $\sH_\bp$ has the usual contributions
from neutrino masses, background matter, and from other neutrinos.

Turning first to the matter effect, we use a local four-fermion
current-current description of weak interactions that is relevant for
the low-energy environments in collapsed stellar cores or in
neutron-star mergers.  This approximation is not sufficient in the
early universe where the background medium is nearly matter-antimatter
symmetric so that the dominant contribution
derives from gauge-boson propagator effects \cite{Notzold:1987ik}.
Moreover, we assume neutral-current interactions to be
independent of flavor, i.e., we ignore radiative corrections
\cite{Botella:1986wy,Mirizzi:2009td}. We finally assume that the
medium is not spin-polarized, but we take into account possible
convective currents. With these assumptions and
working in the weak-interaction basis, the ordinary matter term
depends only on the charged-current contribution of charged
leptons. It takes on the form
$\sH^{\rm matter}_\bp=\sqrt{2}\GF\,v_\alpha\sF_\ell^\alpha$ with the matrix of charged-lepton
fluxes
\begin{equation}
  \sF_\ell^\alpha=\int 2\,d\bp
  \begin{pmatrix}v^\alpha_e\(f_{e,\bp}-\bar f_{e,\bp}\)&0&0\\
    0&v^\alpha_\mu\(f_{\mu,\bp}-\bar f_{\mu,\bp}\)&0\\
    0&0&v^\alpha_\tau\(f_{\tau,\bp}-\bar f_{\tau,\bp}\)
  \end{pmatrix},
\end{equation}
where $\int d\bp=\int d^3\bp/(2\pi)^3$ and
$v^\alpha_e=(1,\bv_e)$ with $\bv_e=\bp/(\bp^2+m_e^2)^{1/2}$,
and similar for $\mu$ and $\tau$ leptons.  Moreover, $f_{e,\bp}$ and $\bar{f}_{e,\bp}$ are the occupation numbers of electrons and positrons with momentum $\bp$, respectively, and analogous for the other charged leptons.  Realistically, of course, in compact stars there are no
$\tau$ leptons, but there will be some population of muons~\cite{Bollig:2017lki}.
If the medium is isotropic, the spatial components vanish and
$\sF_\ell^0$ is simply the matrix of net charged lepton densities
(leptons minus antileptons).

The neutrino-neutrino refractive term has the analogous form
$\sH^{\nu\nu}_\bp=\sqrt{2}\GF\,v_\alpha\sF_\nu^\alpha$ with the
neutrino flux matrix
\begin{equation}\label{eq:nu-flux}
  \sF_\nu^\alpha=\int d\bp\,v^\alpha
  \(\varrho_\bp-\bar\varrho_\bp\)\,,
\end{equation}
where $v^\alpha=(1,\hat\bp)$ and $\varrho_\bp$ and $\bar\varrho_\bp$
are the occupation number matrices for neutrinos and antineutrinos,
respectively. Thus the Hamiltonian matrix is
\begin{equation}\label{eq:EOM3b}
  \sH_{\bp}=\frac{\sM^2}{2E}
  +\sqrt{2}\GF\,v_\alpha\(\sF_\ell+\sF_\nu\)^\alpha\,,
\end{equation}
where $E=|\bp|$. The EOM for the antineutrino matrices $\bar\rho_\bp$ is the same
with a sign change of the vacuum oscillation term. The refractive term
does not depend on $E$, but only on $\bv$. The $E$-independence of the
refractive term owes to the current-current approximation of the
electroweak interaction.

\section{Linearization}
\label{sec:linearization}

\subsection{Small Deviation from Flavor States}

A flavor-dependent neutrino population, for example in a compact
astrophysical object, depends on charged-current production and
interaction processes, typically involving electrons and positron and
perhaps some muons. Therefore, neutrinos are produced in flavor
eigenstates so that the $\varrho$ matrices are diagonal in the flavor
basis and modified by the subsequent effect of flavor conversion. Neutrino
mixing angles are large so that off-diagonal elements of the
$\varrho$ matrices quickly develop in vacuum. However, we are interested in
astrophysical environments where matter effects are large, i.e., the
in-medium effective mixing angles are small and neutrinos remain
essentially pinned to the flavor basis except for the possible effect
of adiabatic conversion (MSW effect) as they pass a resonance
region. We here explicitly exclude such scenarios so that the
off-diagonal $\varrho$ elements remain small unless something new
happens in the form of self-induced flavor conversion caused by
neutrino-neutrino interactions.

In addition, we assume that the system is stationary apart from the
possible effect of neutrino-neutrino interactions, i.e., the
occupation numbers of both neutrinos and charged leptons at a given
location remain constant except for the possible effect of
flavor conversion. To linear order in the small off-diagonal
$\varrho$-elements, the diagonal elements thus remain conserved and we
only ask about the evolution of the off-diagonals which encode flavor
coherence~\cite{Banerjee:2011fj}. Actual flavor conversion, i.e., a non-trivial evolution of
the diagonal $\varrho$-elements, becomes visible only at second order
in the off-diagonals and thus becomes relevant once the evolution
turns nonlinear.

Therefore, we may define an
overall matter effect as an external parameter that is
caused by both charged leptons and neutrinos as
\begin{equation}
  \sH^{\rm matter}=v_\alpha \sLambda^\alpha,
\end{equation}
where $\sLambda^\alpha$ is the diagonal part of
$\sqrt{2}\GF(\sF_\ell+\sF_\nu)^\alpha$. Explicitly
$\sLambda^\alpha={\rm
  diag}(\Lambda_{e}^\alpha,\Lambda_{\mu}^\alpha,\Lambda_{\tau}^\alpha)$
where for each lepton flavor $\ell$
\begin{equation}
  \Lambda_\ell^\alpha=\int d\bp\,
  \Big[v_\ell^\alpha\,2\(f_{\ell,\bp}-\bar f_{\ell,\bp}\)
  +v^\alpha\(f_{\nu_\ell,\bp}-\bar f_{\nu_\ell,\bp}\)\Big]\,.
\end{equation}
Here $v_\ell^\alpha=(1,\bv_\ell)$ with
$\bv_\ell=\bp/(\bp^2+m_\ell^2)^{1/2}$ is the charged-lepton four-velocity
and $v^\alpha=(1,\hat\bp)$ the neutrino four-velocity. In other words,
$\Lambda_\ell^\alpha$ is the four-current of lepton number $\ell$,
carried by both charged leptons and neutrinos.

\subsection{Vanishing Flavor Mixing}

We are aiming at an EOM and mode analysis for the off-diagonal
elements of $\varrho$, an approach that makes strict sense only if
the system has a stationary fixed point in the absence of
neutrino-neutrino interactions. However,
some degree of flavor coherence develops
by the effect of $\sM^2$ alone and indeed in vacuum this effect is
large because neutrino mixing angles are large. On the other
hand, matter effects strongly suppress vacuum flavor conversion, so in
our context, the main effect of the flavor off-diagonal $\sM^2$ elements
will be to trigger instabilities. For the moment we simply take
$\sM^2$ to be diagonal, meaning that we ignore flavor mixing, but not
neutrino masses. In the fast-flavor context, $\sM^2$ was taken to vanish,
whereas here only its off-diagonals are taken to vanish.

In the limit of vanishing neutrino mixing, the linearized EOMs for the
three off-diagonal elements of $\varrho_\bp$ (and their complex conjugates)
decouple, leading to equations of the form
\begin{eqnarray}\label{eq:EOM4}
  i\,v^\alpha\partial_\alpha \varrho_{\bp}^{e\mu}&=&
  \[\frac{\sM^2_{ee}-\sM^2_{\mu\mu}}{2E}
  +v_\alpha(\Lambda_e-\Lambda_\mu)^\alpha\]\varrho_{\bp}^{e\mu}
  \nonumber\\[1ex]
  &&\kern4em{}-\sqrt{2}\GF\(f_{\nu_e,\bp}-f_{\nu_\mu,\bp}\)v^\alpha
  \int d\bp'v_\alpha'\(\varrho_{\bp'}^{e\mu}-\bar\varrho_{\bp'}^{e\mu}\)
\end{eqnarray}
and analogous for the other pairs of flavors.
Therefore, in this approach the three-flavor system corresponds to three
independent two-flavor cases.

Of course, there are three nontrivial cases only if
the distributions of the three flavors are
different. In the supernova context, it was often assumed that
$\nu_\mu$ and $\nu_\tau$ had identical distributions, but
realistically charged muons exist in this environment, implying
non-negligible differences between all flavors.

\subsection{Two-Flavor System}
\label{sec:two-flavor}

The two-flavor EOM further simplifies when we recognize that all flavor
coherence effects depend only on the difference of the original
neutrino distributions. The commutator structure of the EOM implies that
the diagonal parts of all matrices in flavor space drop out. In particular,
we may write the neutrino matrices of occupation numbers in the form
\begin{equation}\label{eq:s-define}
  \varrho_\bp^{e\mu}=\frac{f_{\nu_e,\bp}+f_{\nu_\mu,\bp}}{2}\,\mathbbm{1}
  +\frac{f_{\nu_e,\bp}-f_{\nu_\mu,\bp}}{2}
  \begin{pmatrix}s_\bp&S_\bp\\S_\bp^*&-s_\bp\end{pmatrix}\,,
\end{equation}
where $s_\bp$ is a real number, $S_\bp$ a complex one, and $s_\bp^2+|S_\bp|^2=1$.
To linear order, $s_\bp=1$, so in our linearized system we ask for the
space-time evolution of $S_\bp$ alone which holds all the information concerning
flavor coherence.

Defining the two-flavor matter effect through
$\Lambda^\alpha=(\Lambda_e-\Lambda_\mu)^\alpha$
and the vacuum oscillation frequency
through $\omega_E=(\sM^2_{ee}-\sM^2_{\mu\mu})/(2E)$, the EOM of
equation~\eqref{eq:EOM4} becomes
\begin{equation}\label{eq:EOM5}
  i\,v^\alpha\partial_\alpha S_\bp=
  \bigl(\omega_E+v^\alpha \Lambda_\alpha\bigr) S_{\bp}
  -v^\alpha
  \int d\bp'v_\alpha'\(S_{\bp'}g_{\bp'}-\bar S_{\bp'}\bar g_{\bp'}\).
\end{equation}
An analogous equation applies to the antineutrino flavor coherence
$\bar S_\bp$ with a sign change of~$\omega_E$. Here we use the spectrum
$g_\bp=\sqrt{2}\GF (f_{\nu_e,\bp}-f_{\nu_\mu,\bp})$ and
$\bar g_\bp=\sqrt{2}\GF (f_{\bar\nu_e,\bp}-f_{\bar\nu_\mu,\bp})$,
where we have absorbed $\sqrt{2}\GF$ for notational convenience.

\subsection{Flavor-Isospin Convention}
\label{sec:flavor-isospin}

The structure of these equations becomes both more compact and more
physically transparent in the ``flavor isospin convention'' where we
interpret antiparticles as particles with negative energy and describe
their spectrum with negative occupation numbers. In the
ultrarelativistic limit, neutrino modes are thus described by
$-\infty<E<+\infty$ and their direction of motion $\bv$ with
$\bp=|E|\bv$ and $v^\alpha=(1,\bv)$. The two-flavor spectrum is
\begin{equation}\label{eq:spectrum}
  g_{E,\bv}=\sqrt{2}\GF\begin{cases} f_{\nu_e,\bp}-f_{\nu_\mu,\bp}&\hbox{for $E>0$,}\\
  f_{\bar\nu_\mu,\bp}-f_{\bar\nu_e,\bp}&\hbox{for $E<0$.}
  \end{cases}
\end{equation}
There is no sign-change in the definition of $S$. The EOM thus reads
\begin{equation}\label{eq:EOM6}
  i\,v^\alpha\partial_\alpha S_{E,\bv}=
  \bigl(\omega_E+v^\alpha \Lambda_\alpha\bigr) S_{E,\bv}
  -v^\alpha  \int d\Gamma'\, v_\alpha'\,g_{E',\bv'}  S_{E',\bv'}\,,
\end{equation}
where the phase-space integration is
\begin{equation}
  \int d\Gamma=\int_{-\infty}^{+\infty}\frac{E^2dE}{2\pi^2}\int \frac{d\bv}{4\pi}\,,
\end{equation}
with $\int d\bv$ an integral over the unit surface, i.e., over all
polar angles of $\bp$.  The vacuum oscillation frequency $\omega_E$,
in this convention, automatically changes sign for antineutrinos.  For positive $E$, it is
positive for inverted mass ordering ($\sM_{ee}^2>\sM_{\mu\mu}^2$) and
negative for the normal mass ordering ($\sM_{ee}^2<\sM_{\mu\mu}^2$).

The equations become even more compact if one uses $\omega_E$ as a
parameter to describe the neutrino energy and express the spectrum
$g_{E,\bv}$ instead as $g_{\omega,\bv}$. However, as we here make the
connection to fast flavor conversion which corresponds to $\sM^2\to0$,
it would be cumbersome to take this limit in the $(\omega,\bv)$
language.

\subsection{Normal-Mode Analysis}
\label{sec:mode-analysis}

For a linear EOM it is natural to seek solutions in Fourier space,
i.e., to look for its normal modes and associated dispersion
relation. We thus seek space-time dependent solutions of
equation~\eqref{eq:EOM6} that can be written in the form
\begin{equation}\label{eq:normalmode}
  S_{\Gamma,r}=Q_{\Gamma,K}\,e^{-i(K_0 t-\bK\cdot\br)}\,,
\end{equation}
where $\Gamma=\lbrace E,\bv\rbrace$, $r=(t,\br)$ and $K=(K_0,\bK)$.
The quantity $Q_{\Gamma,K}$ is the eigenvector in $\Gamma$-space
for a given eigenvalue $K$.

To find these normal modes of the EOM we insert the ansatz of
equation~\eqref{eq:normalmode} into equation~\eqref{eq:EOM6} and find
\begin{equation}\label{eq:EOM7}
  \bigl( v_\alpha k^\alpha-\omega_E \bigr) Q_{\Gamma,k}=v_\alpha A^\alpha_k
  \qquad\hbox{with}\qquad
  A^\alpha_k=-\int d\Gamma\, v^\alpha\,g_{\Gamma}  Q_{\Gamma,k}
\end{equation}
and $k^\alpha=K^\alpha-\Lambda^\alpha$. Fully analogous to the fast-flavor case,
we have shifted the original four wavevector,  $K^\mu=(K_0,\bK)$, to the redefined four wavevector, $k^\mu=(k_0,\bk)$, by subtracting the
matter-effect four vector $\Lambda^\mu$. Solving the EOMs in Fourier space is
thus completely independent of the matter effect which has been ``rotated away'' by shifting the origin in the four wavevector space.

In the absence of neutrino-neutrino interactions, the rhs of
equation~\eqref{eq:EOM7} vanishes and nontrivial solutions require
$v_\alpha k^\alpha-\omega_E=0$, i.e., the purely kinematical dispersion
relation $k_0=\omega_E+\bv\cdot\bk$
where each neutrino mode labelled by $\{E,\textbf{v}\}$ evolves independently.
In the presence of neutrino-neutrino
interactions, collective oscillations become possible where this
dispersion relation changes.
Therefore, we consider solutions
with $v_\alpha k^\alpha-\omega_E\not=0$ for all $\{E,\bv\}$
so that equation~\eqref{eq:EOM7} implies
\begin{equation}\label{eq:eigenvector}
  Q_{\Gamma,k}=\frac{v_\alpha A^\alpha_k}{v_\gamma k^\gamma-\omega_E}\,.
\end{equation}
Inserting this form on both sides of equation~\eqref{eq:EOM7} yields
\begin{equation}
  v_\alpha A^\alpha_k=-v^\alpha A^\beta_k\int d\Gamma'\,g_{\Gamma'}\,
  \frac{v'_\alpha v'_\beta}{v'_\gamma k^\gamma-\omega_E}\,.
\end{equation}
In more compact notation this can be written in the form
\begin{equation}\label{eq:EOM8}
  v_\alpha\Pi_k^{\alpha\beta}A_{k,\beta}=0
  \qquad\hbox{with}\qquad
  \Pi_k^{\alpha\beta}=h^{\alpha\beta}+\int d\Gamma\,g_{\Gamma}\,
  \frac{v^\alpha v^\beta}{v_\gamma k^\gamma-\omega_E}\,,
\end{equation}
where $h^{\alpha\beta}={\rm diag}(+,-,-,-)$ is the metric tensor.
This equation must hold for any $v^\alpha$ and thus amounts to
four independent equations $\Pi_k^{\alpha\beta}A_{k,\beta}=0$.
Nontrivial solutions require
\begin{equation}\label{eq:determinant}
  {\rm det}\,\Pi^{\alpha\beta}_k=0\,,
\end{equation}
establishing a connection between the components of $k=(k_0,\bk)$, i.e.,
the dispersion relation of the system. It depends only on the neutrino
flavor spectrum $g(E,\bv)$, which itself contains the neutrino density,
and the energy-dependent vacuum oscillation frequency
$\omega_E$.

Notice that the eigenfunctions $Q_{k}(E,\bv)$ of
equation~\eqref{eq:eigenvector} for collective motions are
not a complete set of solutions of the original EOM. There can be other solutions
which are such that $v_\alpha k^\alpha-\omega_E=0$ for some $\{E,\bv\}$, i.e.,
there exists at least one mode $\{E,\bv\}$ for which the propagation remains
purely kinematical. In the context of our colliding beams model we will
encounter an explicit example of a system that maintains kinematical modes
in the presence of neutrino-neutrino interactions. However, purely kinematical modes will not be unstable, so solving 
equation~\eqref{eq:determinant} is expected to turn up all solutions where
$k_0$ or ${\bf k}$ has an imaginary part.

\subsection{Fast Flavor Limit}

Equation~\eqref{eq:determinant} with the definition of $\Pi^{\alpha\beta}_k$ given
in equation~\eqref{eq:EOM8} is the master equation for all cases of linear
stability analysis performed in the past literature. The case of fast flavor
oscillations corresponds to strictly massless neutrinos so that $\omega_E\to 0$.
In this case the energy integral in equation~\eqref{eq:EOM8} can be performed
on the distribution function alone. We may thus define
\begin{equation}\label{eq:Pi-fastflavor}
  G_\bv=\int_{-\infty}^{+\infty}\frac{E^2 dE}{2\pi^2}\,g_{E,\bv}
  =\sqrt{2}\GF\int_{0}^{\infty}\frac{E^2 dE}{2\pi^2}\,
  \bigl(f_{\nu_e,\bp}-f_{\bar\nu_e,\bp}-f_{\nu_\mu,\bp}+f_{\bar\nu_\mu,\bp}\bigr) \, ,
\end{equation}
with $\bp=E\,\bv$. The $\Pi$ tensor thus becomes
\begin{equation}\label{eq:Pi-fast-flavor}
  \Pi_k^{\alpha\beta}=h^{\alpha\beta}+\int \frac{d\bv}{4\pi}\,G_{\bv}\,
  \frac{v^\alpha v^\beta}{v_\gamma k^\gamma} \, ,
\end{equation}
in agreement with earlier results~\cite{Izaguirre:2016gsx}.

In the fast flavor limit of $\omega\to0$ the EOM is the same for
all modes $S_{E,\bv}$ with equal $\bv$ but different $E$. In
reference~\cite{Izaguirre:2016gsx} the discussion was formulated
as if one only needed an EOM for $S_{\bv}$, taken to be the same
for all values of $E$. However, this assumption is not justified
because the different modes $S_{E,\bv}$ need not evolve the same
for all $E$ just because they obey the same EOM---their evolution
also depends on the initial conditions that could be different
for different $E$, depending on what triggers a nontrivial evolution.

\subsection{Multi-Angle Matter Effect}

The discussion so far appears to imply that background matter plays no
role in collective flavor conversion because it was removed by the
shift $K^\alpha\to k^\alpha=K^\alpha-\Lambda^\alpha$. On the other
hand, in the previous literature it appeared that the multi-angle
matter effect would suppress collective flavor conversion in many
cases of practical astrophysical interest. This apparent discrepancy
is resolved if we remember that in much of the past literature it was
assumed that a stationary astrophysical background model implied
time-independent solutions for collective flavor
conversion. Therefore, only time-independent solutions were
considered, which in our terminology implies $K_0=0$. With this
assumption, the polarization tensor is
\begin{equation}
  \Pi_k^{\alpha\beta}=h^{\alpha\beta}-\int d\Gamma\,g_{\Gamma}\,
  \frac{v^\alpha v^\beta}{\bv\cdot\bK+v_\gamma\Lambda^\gamma+\omega_E}\,.
\end{equation}
On this basis we can look for eigenvalues $\bK$ and their imaginary
parts. This will lead to the same results as those found in
reference~\cite{Chakraborty:2015tfa}.

However, solutions which are not stationary in this sense may have
much larger growth rates. Therefore, the role of the matter effect
depends on the temporal behavior of the solution \cite{Abbar:2015fwa,
  Dasgupta:2015iia,Capozzi:2016oyk}. Therefore, it is crucial to look at collective
flavor conversion as a space-time dependent effect, not just one that
by fiat depends only on time alone or only on space alone.

\subsection{Discrete Modes}

The derivation of the dispersion relation of
equation~\eqref{eq:determinant} leaves a number of questions open,
in particular concerning the completeness of
solutions. An alternative approach, in particular for numerical
studies, is to consider a discrete mode spectrum
$g_j=g_{E_j,\bv_j}$ with \hbox{$j=1,\ldots, N$}
so that the phase-space integral $\int dE\,d\bv$ turns into
a summation. The initial EOM of equation~\eqref{eq:EOM6}
is a set of $N$ equations and the normal-modes $Q_{j,k}$
become \hbox{$N$-dimensional} vectors of complex numbers
fulfilling the equation
\begin{equation}
  \sum_{i=1}^{N}\Bigl[\(v_{j,\alpha} k^\alpha-\omega_j\)\delta_{ij}
   +v_{j,\alpha}
  v_i^\alpha g_i\Bigr] Q_{i,k}=\sum_{i=1}^{N} M_{ij,k} Q_{i,k}=0\,.
\end{equation}
The dispersion relation follows from the condition
\begin{equation}\label{eq:discrete-determinant}
  D(k_0,\bk)=
  {\rm det}\,M_k=0\,.
\end{equation}
This is a purely polynomial equation and thus provides
us with $N$ solutions. In this approach one need not divide
by the potentially vanishing term $(v_{j,\alpha} k^\alpha-\omega_j)$
and thus one will not miss any solutions.
The condition of equation~\eqref{eq:discrete-determinant}
is equivalent to equation~\eqref{eq:determinant} for the non-singular
cases owing to the degeneracy (separability) of
the ``kernel'' of the summation (or integral) equation presently at
hand.

Typically a discrete neutrino distribution $g_j$ will be used, for example
in numerical
studies, to represent a continuous distribution $g_{E,\bv}$.
The virtue of finding a complete set of solutions in the discrete case
can now become the vice of ``too many'' solutions which have been
dubbed ``spurious modes'' \cite{Sarikas:2012ad, Morinaga:2018aug}.
Of course, they are spurious only in the sense
that they may not appear in the continuous case that we wish to represent.
Recently it was shown in a specific example that in the limit $N\to\infty$ the spurious solutions of $D(k_0,\bk)=0$ merge to form a branch
cut of this complex function \cite{Morinaga:2018aug}.
However, it is not assured that this will always be the case, i.e.,
conceivably the discrete-mode approach in the $N\to\infty$ limit
might reveal ``non-spurious'' solutions that were missed by
the ${\rm det}\, \Pi_k^{\alpha\beta}=0$ criteria,
although no such example appears to exist in the literature.

\section{Slow and Fast Modes}
\label{sec:slowfast}

In order to illustrate existence of slow and fast modes,  the
simultaneous action of slow and fast modes, and later to
study the role of flavor mixing to trigger instabilities,
we set up a simple model that has proven useful
as a theoretical laboratory.

\subsection{Colliding Beams Model}
\label{sec:two-beam}

We consider a homogeneous system that is essentially
one-dimensional, i.e., all neutrino modes are along or opposite the
$z$-direction (``colliding beams''). Moreover, we only consider
solutions that vary in time and/or in the $z$-direction. We use
altogether 4~modes, with Nos.~1 and~3 having $v_z=+1$ and Nos.~2 and~4
having $v_z=-1$. For the moment we leave the four energies $E_j$
($j=1,\dots,4$) unspecified, giving us four vacuum oscillation
frequencies $\omega_j=\Delta m^2/(2E_j)$. We work in the flavor isospin
convention, so negative energies will signify antineutrinos.  The EOM
for the four complex flavor coherence functions $S_j(t,z)$ following
from equation~\eqref{eq:EOM6} are then explicitly
\begin{subequations}\label{eq:EOM20}
  \begin{eqnarray}
    i(\partial_t+\partial_z)S_1 &=&(\omega_1+\Lambda_0-\Lambda_z)S_1-\mu_2S_2-\mu_4S_4,\\
    i(\partial_t-\partial_z)S_2 &=&(\omega_2+\Lambda_0+\Lambda_z)S_2-\mu_1S_1-\mu_3S_3,\\
    i(\partial_t+\partial_z)S_3 &=&(\omega_3+\Lambda_0-\Lambda_z)S_3-\mu_2S_2-\mu_4S_4,\\
    i(\partial_t-\partial_z)S_4 &=&(\omega_4+\Lambda_0+\Lambda_z)S_4-\mu_1S_1-\mu_3S_3.
  \end{eqnarray}
\end{subequations}
The neutrino-neutrino interaction potentials are
$\mu_j=2\sqrt{2}\GF(n_j^{\nu_e}-n_j^{\nu_\mu})$ with
$n_j^{\nu_e,\nu_\mu}$ the number densities of flavor $\nu_e$ or
$\nu_\mu$ in mode $j$. For antineutrino modes (depending on the sign
of $E_j$), we have instead
$\mu_j=2\sqrt{2}\GF(n_j^{\bar\nu_\mu}-n_j^{\bar\nu_e})$. The factor 2
arises from $v^\alpha v'_\alpha$ of the two modes, which is 0 for
parallel and 2 for anti-parallel velocities.
In this way in equation~\eqref{eq:EOM20} only the
even-numbered modes influence the odd-numbered ones and the other way
around. The neutrino part of the matter potential is
$\Lambda_{\nu,0}=(\mu_1+\mu_2+\mu_3+\mu_4)/2$ and
$\Lambda_{\nu,z}=(\mu_1-\mu_2+\mu_3-\mu_4)/2$.

In order to cover all simple examples we
consider a monochromatic system, meaning that $|E_j|$ is the same for
all modes. We take modes No.~1 and No.~4 to consist of neutrinos, so we
use $\omega\equiv\omega_1=\omega_4$, and Nos.~2 and~3 to consist of
antineutrinos, so $\omega_2=\omega_3=-\omega$, as shown in
table~\ref{tab:mode-occupation}. Moreover, for inverted mass ordering
we have $\omega>0$, otherwise $\omega<0$, i.e., we define
$\Delta m^2=m_1^2-m_2^2$ to be positive for inverted mass ordering.

\begin{table}[ht]
  \centering
\begin{tabular}{cccccc}
  \hline
  Mode No.& Particle  & $\omega_j$ & Direction & $v_z$ & $\mu_j$\\
  \hline\hline
  1 & $\nu$     & $+\omega$ & $\rightarrow$ & $+1$ & $+g_1\mu$ \\
  2 & $\bar\nu$ & $-\omega$ & $\leftarrow$  & $-1$ & $-g_2\mu$ \\
  3 & $\bar\nu$ & $-\omega$ & $\rightarrow$ & $+1$ & $-g_3\mu$ \\
  4 & $\nu$     & $+\omega$ & $\leftarrow$  & $-1$ & $+g_4\mu$ \\
  \hline
\end{tabular}
\caption{Properties of the four modes in our colliding beams model.
  Here $\omega=\Delta m^2/2|E|$ is positive for inverted mass
  ordering and negative otherwise.
  The mode occupations $g_j$ are positive
  numbers if there is an excess of $\nu_e$ and $\bar{\nu}_e$ compared to $\nu_\mu$ and $\bar{\nu}_\mu$, respectively, for all modes.}
\label{tab:mode-occupation}
\end{table}

We also introduce dimensionless mode occupations $g_j$ such that
$\mu_j=g_j\mu$ with the common positive scale factor,
\begin{equation}
  \mu=\frac{2\sqrt{2}\GF}{4}\biggl(\sum_{j=1,4}|n_j^{\nu_e}-n_j^{\nu_\mu}|+
  \sum_{j=2,3}|n_j^{\bar\nu_e}-n_j^{\bar\nu_\mu}|\biggr).
\end{equation}
Separating the overall neutrino-neutrino interaction strength $\mu$ from the
``spectrum'' $g_j$ is not unique, but it is instructive to separate
the overall scale from the detailed relative mode occupations. The normalization condition is chosen as $\sum_{j=1}^4|g_j|=4$.
If we assume an initial excess
of electron-flavored neutrinos and antineutrinos over their muon-flavored counterparts,
the mode occupations $g_j$ are positive with $\sum_{j=1}^4g_j=4$.

The diversity and complexity of collective neutrino oscillations derives
from the dependence on initial conditions encoded in the mode occupations $g_i$. Different choices of $g_i$ lead to qualitatively different evolutions.
To illustrate this point, we will first consider that all mode occupations are
equal, say $g_i=g$. We dub this highly symmetric case as the
``flavor pendulum.'' Then we will consider the more generic case, where the various $g_i$ are not all equal, which we will refer to as the general colliding beams model.

\subsection{Dispersion Relation and Classification of Modes}

We look for solutions of the form $S_j(t,z)=Q_je^{-i(K_0t-K_z z)}$,
leading to
$(K_0-K_z)Q_1=(\omega_1+\Lambda_0-\Lambda_z)Q_1-\mu_2Q_2-\mu_4Q_4$ for
the first equation, and analogous for the others. We shift the
wave vector by virtue of $(k_0,k_z)=(K_0,K_z)-(\Lambda_0,\Lambda_z)$,
so our equations become
\begin{equation}\label{eq:EOM:21}
  \begin{pmatrix}
    (\omega-k_0+k_z)&g_2\mu&0&-g_4\mu\\
    -g_1\mu&(-\omega-k_0-k_z)&g_3\mu&0\\
    0&g_2\mu&(-\omega-k_0+k_z)&-g_4\mu\\
    -g_1\mu&0&g_3\mu&(\omega-k_0-k_z)\\
  \end{pmatrix}
  \begin{pmatrix}
    Q_1 \\
    Q_2 \\
    Q_3 \\
    Q_4
  \end{pmatrix}=0\,.
\end{equation}
Non-trivial solutions, and hence the dispersion relation, follow from
the requirement that the determinant of the
$4{\times}4$ matrix in this equation vanishes.

The dispersion relation for our colliding beams
system following from the condition that the determinant of the matrix in
equation~\eqref{eq:EOM:21} vanishes is given by
the general expression
\begin{eqnarray}\label{eq:Det0}
  \(k_0^2-k_z^2\)\[k_0^2-k_z^2+(g_1-g_3)(g_2-g_4)\mu^2\]
  &=&2\[(g_1g_4-g_2g_3)k_0+(g_3g_4-g_1g_2)k_z\] \mu^2\omega
  \nonumber\\[1ex]
  &+&\[2\(k_0^2+k_z^2\)+(g_1+g_3)(g_2+g_4)\mu^2\]\omega^2
  \nonumber\\[1ex]
  &-&\omega^4.
\end{eqnarray}
A general solution of this quartic equation is complicated
and not informative, but we can explicitly consider several limiting
cases.

Unless each of the two colliding beams consists of only neutrinos or
only antineutrinos, a special case that one may consider, each beam
consists of two modes. Therefore, generally there are four nontrivial
solutions, corresponding to the four modes of our system. These modes
typically have different symmetries and it is a different question if
they get equally excited by a given perturbation in flavor space that
may favor one symmetry over another.  To address these questions one
also needs to study the properties of eigenvectors corresponding to
the four eigenvalues.

In the limit $\mu\to0$ where neutrinos do not interact with each
other, we get four branches of the dispersion relation
$k_0=\pm k_z\pm\omega$. These solutions are not dynamical in that
they correspond to a perturbation on a given mode to simply drift
along with the velocity of that mode. We call these ``inert'' or
``kinematical'' solutions.

For nonzero $\mu$ one typically finds ``dynamical'' modes. These are solutions
that do not correspond to a perturbation simply drifting along the velocity.
Further, if a mode is dynamical in the limit $\omega\to0$, we call it a
``fast mode,'' and otherwise a ``slow mode.'' Obviously, the latter are
inert in the limit that neutrinos are massless.

We are ultimately interested in modes that can grow
exponentially with time or space. The dispersion relation for
dynamical propagating modes, i.e., those with both $k_0$ and
$k_z$ being real, do not cross, so typically
there will be ``forbidden'' ranges of $k_z$ where $k_0$ has an
imaginary part or the other way round. These can be classified as absolute or
convective instabilities \cite{Capozzi:2017gqd}, with
the imaginary parts determining the growth rate.  If the growth rate does
not depend on $\omega$, but requires only a nonvanishing $\mu$, we say that the mode has a ``fast instability.'' If a nonvanishing $\omega$ is required, so that the growth
rate vanishes in the $\omega\to0$ limit, we say that the mode has a
``slow instability.'' Note that fast modes can have fast or slow
instabilities, whereas slow modes can only have slow instabilities. The former happens when fast and slow modes mix with each other, as demonstrated explicitly in section~\ref{sec:fast-slow-modes}.

In our colliding beams model, fast modes are the solutions which arise when
the rhs of equation~\eqref{eq:Det0} vanishes. In this limit of $\omega \to0$ there
exist two purely kinematical solutions with $k_0 = \pm k_z$ and two dynamical ones
$k_0 = \pm \sqrt{k_z^2 + m^2}$, where $m^2 = (g_1 - g_3 )(g_4 - g_2) \mu^2$.
The latter are either analogous to a particle with mass $|m|$ or a tachyon with
imaginary mass $i|m|$, depending on the sign of $(g_1 - g_3)(g_4 - g_2)$. For
tachyonic solutions one finds imaginary $k_0$, signaling an instability.
In this case, the group velocity $dk_0 / dk_z$ is superluminal as opposed
to the particle-like solutions. However, this behavior does not imply a violation of
causality because signals still propagate only within the light cone, i.e., the
group velocity is not a reliable measure for signal propagation.  The presence
of imaginary frequencies for some range of real wave numbers prevents a
localized perturbation to propagate with
superluminal speed \cite{Aharonov:1969vu}. Other modes require $\omega\not=0$
to be dynamical and their possible instabilities have the same requirement. From
linear perturbation for small $\omega$ one can easily show that in this case the
growth rate scales with $\sqrt{|\omega|}$, which is characteristic for slow instabilities.

\subsection{Modes in the Colliding Beams Model}

Now we will explore how different modes can arise in our
colliding beams model, depending on the values of $g_i$ and $\omega$.
In particular, we will see that it is possible that a solution that is inert
in the limit $\omega\to0$, after including a non-vanishing $\omega$
becomes dynamical and acquires a slow instability.

\subsubsection{Only Slow Modes for the Flavor Pendulum}
\label{sec:slow-modes-only}

\begin{figure}[!t]
  \centering
  \includegraphics[width=0.7\textwidth]{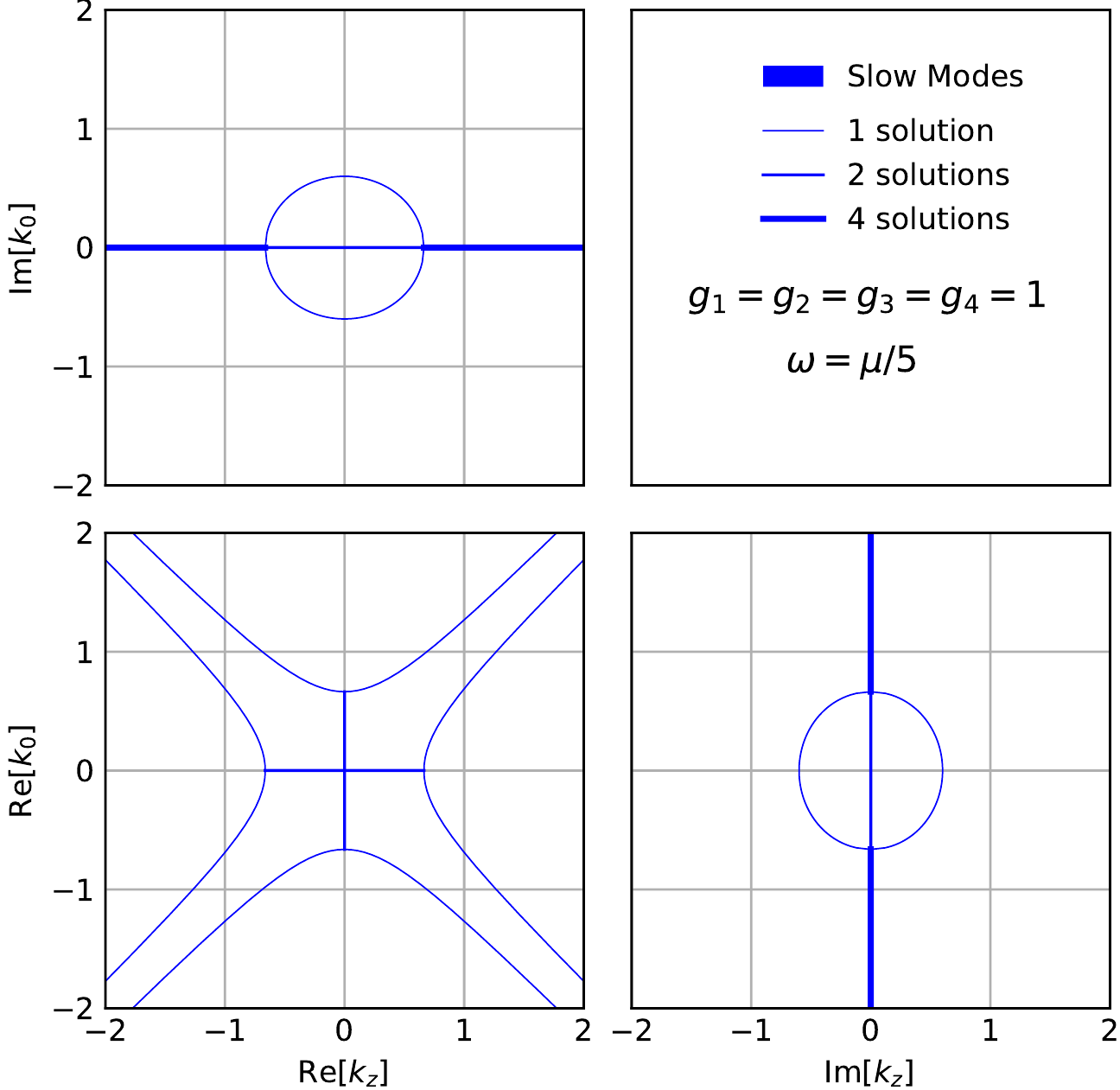}
  \caption{Dispersion relation of equation~\eqref{eq:dispersion1} for our isotropic colliding beams model with equal $\nu$ and $\bar\nu$ densities.
  This choice of $g$'s is the only case for which we have slow modes only, i.e., in the limit $\omega \rightarrow 0$ all modes are non-dynamical.
  All frequencies are shown in units of $\mu$.
  This figure can be read with real $k_z$ on the horizontal axis as independent variable, showing ${\rm Re}(k_0)$
  in the lower and ${\rm Im}(k_0)$ in the upper panel. Or we can use
  real $k_0$ on the vertical axis as independent variable and show
  the corresponding ${\rm Re}(k_z)$ in the lower left panel and
  ${\rm Im}(k_z)$ on the lower-right panel.}
  \label{fig:slowdispersion}
\end{figure}

The flavor pendulum, i.e., an isotropic gas of equal $\nu$ and $\bar\nu$
densities corresponds in our 1D model to $g_1=g_2=g_3=g_4=1$, leading
to the explicit dispersion relation
\begin{equation}\label{eq:dispersion1}
  k_0^2=\omega^2+k_z^2\pm2\sqrt{\omega^2\bigl(k_z^2+\mu^2\bigr)}\,.
\end{equation} Obviously they are all slow modes.
We show these solutions for propagating modes (real $k_0$ vs.
real $k_z$) in the lower left panel of
figure~\ref{fig:slowdispersion}. There are four solutions.
One pair of branches shows a
frequency gap when for real $k_z$ we get $k_0$ with nonzero imaginary part.
The other pair shows a gap in wavenumbers.  So there are
``forbidden'' bands with imaginary frequencies or wavenumbers, i.e.,
possibly unstable solutions. These instabilities can be further classified as
absolute vs.\ convective following reference~\cite{Capozzi:2017gqd}
for fast modes.
We show the imaginary parts, relevant in the gaps, in the top and right panels.

It is instructive to consider specifically the homogeneous solutions
($k_z=0$) and the corresponding eigenvectors. We may write and group
them as
\begin{subequations}\label{eq:eigenvectors}
\begin{eqnarray}
  \kern-2em\hbox to 4em{Even:\hfil}
  k_0=\pm\sqrt{\omega^2-2\mu\omega}
  &\quad\hbox{and}\quad&
  Q\propto\begin{pmatrix}
      \mu \\
      \mu-\omega\pm\sqrt{\omega^2-2\mu\omega}\\
      \mu-\omega\pm\sqrt{\omega^2-2\mu\omega} \\
      \mu
    \end{pmatrix},
\\
  \kern-2em\hbox to 4em{Odd:\hfil}
  k_0=\pm\sqrt{\omega^2+2\mu\omega}
  &\quad\hbox{and}\quad&
  Q\propto\begin{pmatrix}
      -\mu \\
      \mu+\omega\mp\sqrt{\omega^2+2\mu\omega}\\
      -\bigl(\mu+\omega\mp\sqrt{\omega^2+2\mu\omega}\bigr) \\
      \mu
    \end{pmatrix}.
\end{eqnarray}
\end{subequations}
Recall that the eigenvector $Q$
has its components ordered by the mode numbers in table~\ref{tab:mode-occupation}.
For the even solutions, neutrinos of the two opposing beams contained in mode nos. 1 and 4 behave the same,
as do antineutrinos of the two beams contained in mode nos. 2 and 3. In the odd solutions, the same particles
of opposing beams have opposite amplitudes.
Assuming $\mu>|\omega|/2$ and for $\omega>0$ (inverted mass ordering)
the even solution has an imaginary $k_0$ and thus is unstable, whereas
the odd solution has an oscillating one.
For $\omega<0$ (normal ordering), it is the odd solution that is
unstable. Therefore, the different branches in figure~\ref{fig:slowdispersion}
can be classified according to these symmetry properties that apply at $k_z=0$.

\subsubsection{Fast Modes}
\label{sec:fast-modes}
\begin{figure}[!b]
	\centering
	\includegraphics[width = 0.7\textwidth]{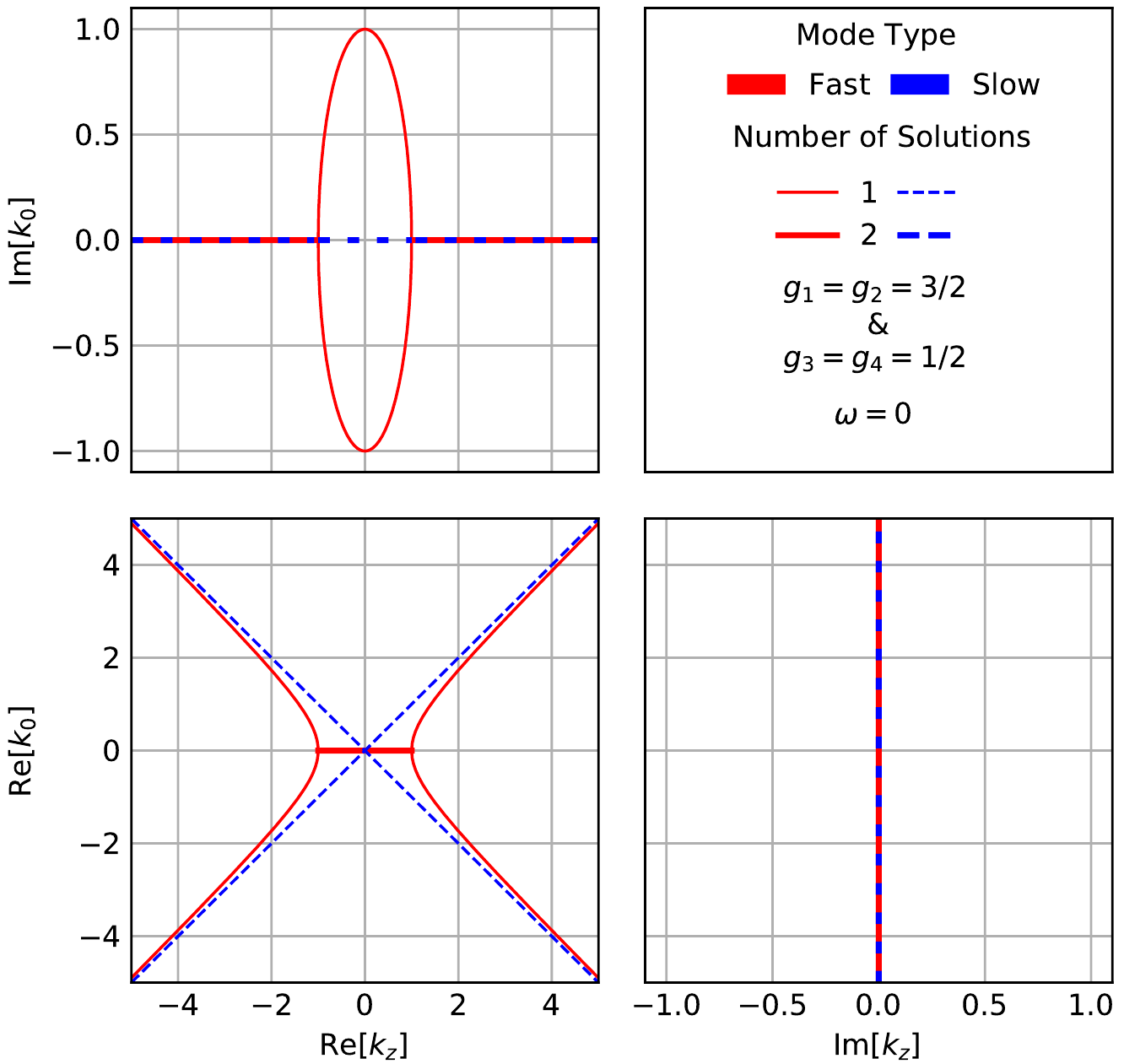}
	\caption{Dispersion relation of equation~\eqref{eq:Det1}. There are both fast modes (continuous lines)
	and slow modes (dashed lines) in the limit of vanishing $\omega$. The fast modes exhibit an instability for
	small values of $k_z$, as evident from the upper left panel.
	The real and imaginary parts are understood as in figure~\ref{fig:slowdispersion}, with all frequencies shown in units of $\mu$.
	Note the change of scale for the imaginary axes.}
	\label{fig:fast_only}
\end{figure}

In general, with generic values of $g_i$ one finds both fast and slow modes.
We recall that by ``fast modes'' we mean those showing dynamical behavior in the
$\omega\to 0$ limit. The general dispersion relation of equation~\eqref{eq:Det0}
reads in this case
\begin{equation}\label{eq:Det1}
  \(k_0^2-k_z^2\)\(k_0^2-k_z^2-m^2\)=0
  \quad\hbox{where}\quad
  m^2=(g_1-g_3)(g_4-g_2)\mu^2.
\end{equation}
Figure~\ref{fig:fast_only} shows representative solutions for the tachyonic case, i.e.,
$m^2=(g_1-g_3)(g_4-g_2)<0$. Two solutions (dashed lines) are inert,
while the other two correspond to dynamical fast modes with a fast instability. Note that the neutrino mass ordering is irrelevant in this limit.

The net occupation $g_+\equiv g_1-g_3$ is proportional to the number
densities of $\nu_e-\bar\nu_e-\nu_\mu+\bar\nu_\mu$ in the $v_z=+1$
beam. If initially the $\nu_\mu$ and $\bar\nu_\mu$ densities are the
same, this is the electron lepton number (ELN) carried by the $v_z=+1$
beam. Likewise, $g_-\equiv g_4-g_2$ is the ELN carried by the $v_z=-1$
beam. Therefore, the dispersion relation is particle-like for $g_+
g_->0$, i.e., both beams carry an excess of ELN. In the opposite
situation with ``a crossing of the angle-spectrum,'' i.e., when $g_+
g_-<0$ the dispersion relation is tachyon-like. In this case the
system shows exponential growth in time for some range of
wavenumbers.

The eigenvectors of these four modes can be explicitly determined and
written in the following form
\begin{subequations}\label{eq:eigenvectors-fast}
\begin{eqnarray}
  k_0=+k_z
  &\quad\hbox{and}\quad&
  Q\propto\begin{pmatrix}
      g_3\\
      0\\
      g_1\\
      0
    \end{pmatrix}, \label{eq:slow-eigenvector1}
\\[2ex]
  k_0=-k_z
  &\quad\hbox{and}\quad&
  Q\propto\begin{pmatrix}
      0 \\
      g_4\\
      0\\
      g_2
  \end{pmatrix},\label{eq:slow-eigenvector2}
\\[2ex]
  k_0=\pm\sqrt{k_z^2+m^2}
  &\quad\hbox{and}\quad&
  Q\propto\begin{pmatrix}
      \mp\sqrt{k_z^2+m^2}-k_z \\[1ex]
      (g_1-g_3)\mu\\[1ex]
      \mp\sqrt{k_z^2+m^2}-k_z \\[1ex]
      (g_1-g_3)\mu
    \end{pmatrix}.\label{eq:fast-dyn-eigenvectors}
\end{eqnarray}
\end{subequations}
The first two are inert solutions, corresponding to the eigenvalues shown using
dashed lines in figure~\ref{fig:fast_only}, where the neutrinos and
antineutrinos of one beam have amplitudes such that the last two terms exactly
cancel in the rhs of equation~\eqref{eq:EOM20} and do not influence the opposite-moving beam.
The dynamical modes, on the other hand, are such that neutrinos and antineutrinos of
a given beam can indeed be grouped together and are described by the
same coherence function. Notice however that these eigenvectors,
while being linearly independent, are not mutually orthogonal except for the two inert modes.

\subsubsection{Mixing of Slow and Fast Modes}
\label{sec:fast-slow-modes}

\begin{figure}[!t]
  \centering
  \includegraphics[width=0.7\textwidth]{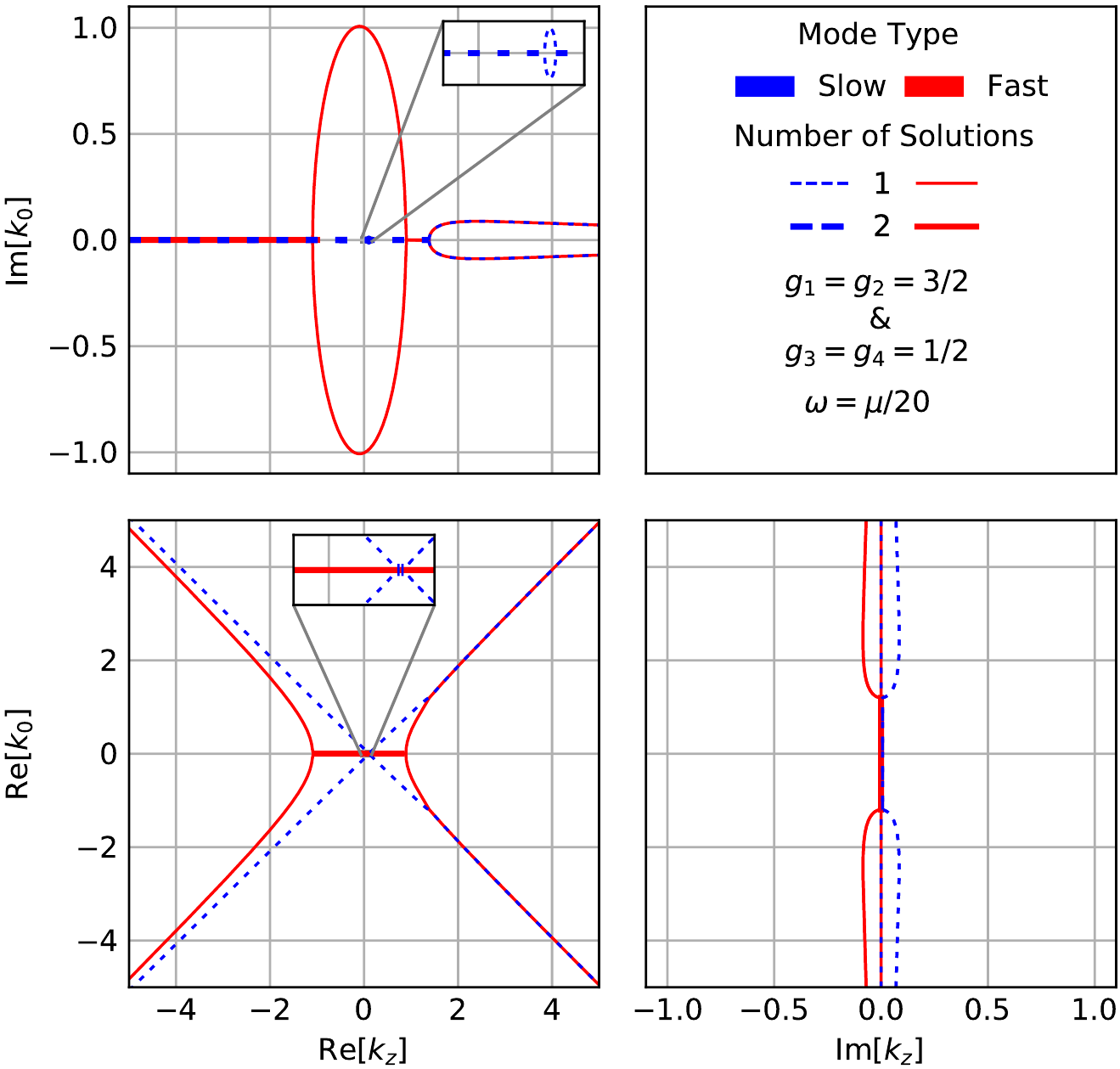}
  \caption{Dispersion relation following from equation~\eqref{eq:Det0}. Slow modes (dashed lines) that were inert for
  vanishing $\omega$ now become dynamical. This is clear not only near the origin, as evident from
  the inset in the upper right panel, but also at large $k_z$ due to mixing between fast and slow modes.  Such a behavior is
  generic except for some very special parameter choices.  The real and imaginary parts are understood as in
    figure~\ref{fig:slowdispersion} with all frequencies shown in units of $\mu$.}
  \label{fig:fastdispersion}
\end{figure}

Next we include nonvanishing neutrino masses and thus take
$\omega\not=0$. Now the previously inert modes also become dynamical
and, more generally, the previous four solutions get mixed with
each other. In figure~\ref{fig:fastdispersion} we show an example for
the dispersion relation, corresponding to the same model parameters as the previous section, but now including a small nonzero $\omega>0$.

Near the crossing region of the previously inert modes when at the origin,
where $|k_z|<|\omega|$, these modes become dynamical and for
the chosen parameters develop a gap in $k_z$, i.e., they are
tachyonic in this region (see the inset in figure~\ref{fig:fastdispersion}).
We can see this behavior analytically starting from the general dispersion relation in equation~\eqref{eq:Det0} and finding the corrections to inert modes
introduced due to non-zero $\omega$. After a first perturbative
iteration, keeping terms up to quadratic powers of $\omega$ and $k_z$, the dispersion relation are
\begin{equation}
	k_0^2 \simeq k_z^2 + \frac{2(g_3g_4 - g_1g_2)}{(g_1-g_3)(g_2-g_4)}\omega k_z + \frac{(g_1+g_3)(g_2+g_4)}{(g_1-g_3)(g_2-g_4)}\omega^2 - \frac{(k_z^2-\omega^2)^2}{(g_1-g_3)(g_2-g_4)\mu^2}\,.
\end{equation}

For sufficiently large $|k_z|$ the inert modes $k_0=\pm k_z$ and the dynamical ones
$k_0=\pm\sqrt{k_z^2+m^2}$ get so close to each other that the
small perturbation provided by non-vanishing $\omega$ can be enough to mix them strongly.
Quantitatively, this happens for
\begin{equation}\label{cond1}
|k_z| \gtrsim \left|\frac{(g_1 - g_3)^2 (g_2 - g_4) \mu^2}{8(g_1+g_3)\omega}\right|,
\end{equation}
providing the roots after mixing
\begin{equation}
	k_0 \simeq k_z \pm \sqrt{ \omega^2 - \frac{(g_1+g_3)(g_2 - g_4)\omega \mu^2}{2k_z}}\,.
\end{equation}
In our example, $(g_1+g_3)(g_2-g_4)>0$ so that for $\omega>0$ and provided that equation~\eqref{cond1} holds true, the square-root becomes imaginary for a range of values of $k_z$ (it stays real for $k_z \rightarrow \infty$), leading to the
non-vanishing ${\rm Im}(k_0)$ shown in the upper panel of figure~\ref{fig:fastdispersion}.
Likewise, for a chosen real $k_0$ in this region, $k_z$ develops an imaginary part.
For negative $k_z$, the modes do not mix but get pushed apart, never becoming asymptotically
close. Changing the sign of $\omega$ is equivalent to changing the sign of
$k_z$, i.e., depending on the sign of $\omega$ (the mass ordering) this
plot flips horizontally.

\subsection{Section Summary}

Except for very special choices of parameters that preserve certain symmetries,
e.g., as in the isotropic limit of our colliding beams model discussed in section \ref{sec:slow-modes-only},
one generically finds fast collective flavor modes, i.e., they are dynamical even for vanishing neutrino
masses when $\omega=0$. In keeping with previous findings, the dispersion relation is tachyonic
if the beams carry opposite ELN, otherwise it is particle-like.

Including neutrino masses in the form of non-vanishing $\omega$ would naively seem to
suggest only a small correction. However, in addition to the fast dynamical modes there exist
inert ones that had been ignored in previous two-beam
studies \cite{Izaguirre:2016gsx, Capozzi:2017gqd}. When these additional
modes cross each other or when they become asymptotically close to the fast dynamical ones,
even a small $\omega$ is enough to mix them and can provide new instabilities. These new instabilities
however are slow, i.e., possible growth rates scale with $\sqrt{|\omega/\mu|}$ compared
with fast instabilities.

The two-beam examples in references \cite{Izaguirre:2016gsx, Capozzi:2017gqd} were
constructed somewhat differently in that the two ``beams'' represented
two different cones of azimuth-integrated modes with different zenith angle
streaming from a source. In this case the velocities represent the radial motion
from the source and as such are less than the speed of light. Further,
the two inert modes and the two fast dynamical ones have different asymptotes
for large $k_z$, i.e., they do not get asymptotically close and do not get mixed
by a small $\omega$. However, there are several intersection points, providing
``bubbles'' of slow instabilities similar to what happens near the origin
of figure~\ref{fig:fastdispersion}.

\section{Triggering of Modes}
\label{sec:flavor-mixing}

In this section we discuss how the flavor mixing term in the vacuum Hamiltonian and non-uniform
matter density are responsible for exciting the various normal-modes we found using the dispersion relation.

\subsection{Flavor Mixing}

Thus far our discussion has assumed the absence of flavor mixing so that
the mass matrix was diagonal in the weak-interaction basis. In this way,
diagonal $\varrho$ matrices represent a fixed point of the EOM and small
perturbations in space and time lead to stable or unstable wave solutions.
Of course, in the absence of any source of flavor violation,
no stable or unstable perturbations can be excited except perhaps by
quantum fluctuations. On the other hand, the observed $\sM^2$ matrix is
far from diagonal and it is the only known source of leptonic flavor violation.
We here always ignore hypothetical flavor-violating non-standard interactions.

Even though the off-diagonal $\sM^2$ elements are large, initially the
off-diagonal $\varrho$ elements are small so that we may still linearize
the EOM. Equation~\eqref{eq:EOM4}, for the off-diagonal element $\varrho^{e\mu}$,
then acquires the following additional terms on the rhs
\begin{equation}\label{eq:newterms}
  -\frac{\sM^2_{e\mu}}{2E}\,\bigl(\varrho^{ee}_\bp-\varrho^{\mu\mu}_\bp\bigr)
  +\frac{\sM^2_{e\tau}}{2E}\,\varrho^{\tau\mu}_\bp
  -\frac{\sM^2_{\tau\mu}}{2E}\,\varrho^{e\tau}_\bp\,.
\end{equation}
Therefore, the linearized three-flavor EOMs
for the three coherences $\varrho^{e\mu}$, $\varrho^{\mu \tau}$ and
$\varrho^{\tau e}$ no longer separate as independent
two-flavor systems --- flavor coherence among any two flavors is communicated to
the others. We here do not pursue such three-flavor effects any further and
return to a two-flavor discussion.

In a two-flavor system, for which we symbolically use the $e\mu$ system, the
relevant components of the matrix of squared neutrino masses are
\begin{subequations}
\begin{eqnarray}\label{eq:two-flavor-masses}
  \omega_E^{\rm c}&=&\frac{\sM_{ee}^2-\sM_{\mu\mu}^2}{2E}
  =\frac{\Delta m^2}{2E}\,\cos2\theta\,,
  \\[2ex]
   \omega_E^{\rm s}&=&\kern1.6em\frac{\sM_{e\mu}^2}{2E}\kern1.6em
  =\frac{\Delta m^2}{2E}\,\sin2\theta\,,
\end{eqnarray}
\end{subequations}
where $\Delta m^2=m_1^2-m_2^2$ is the difference of the squared mass eigenvalues
and $\theta$ is the mixing angle. These frequencies are positive for positive $E$
and for inverted mass ordering.
The EOM in flavor-isospin convention of equation~\eqref{eq:EOM6} thus becomes
\begin{equation}\label{eq:EOM10}
  i\,v^\alpha\partial_\alpha S_{E,\bv}=-\omega_E^{\rm s}+
  \bigl(\omega_E^{\rm c}+v^\alpha \Lambda_\alpha\bigr) S_{E,\bv}
  -v^\alpha  \int d\Gamma'\, v_\alpha'\,g_{E',\bv'}  S_{E',\bv'}\,,
\end{equation}
and thus has acquired a constant term, i.e., the first term on the rhs. Initially $S_{E,\bv}=0$,
so it is this new term which starts any possible motion of the system.

\subsubsection{Flavor Pendulum}

As an explicit example of triggering of modes, we consider the
model of section~\ref{sec:two-beam} in the isotropic
limit in one dimension, i.e., the number of neutrinos and antineutrinos going in opposite directions are equal ($g_1 = g_4\text{ and }g_2= g_3$), and with equal densities of neutrinos and antineutrinos,
i.e., the flavor pendulum where only slow
modes appear. In this case the EOM is
\begin{equation}\label{eq:EOM40}
  i\partial_t\begin{pmatrix}
               S_1 \\
               S_2 \\
               S_3 \\
               S_4
             \end{pmatrix}=
   \begin{pmatrix}
               -\omega^{\rm s} \\
               \omega^{\rm s} \\
               \omega^{\rm s} \\
               -\omega^{\rm s}
   \end{pmatrix}+
    \begin{pmatrix}
               \Lambda_0{+}\omega^{\rm c}{-}\Lambda_z&\mu&0&-\mu\\
               -\mu&\Lambda_0{-}\omega^{\rm c}{+}\Lambda_z&\mu&0\\
               0&\mu&\Lambda_0{-}\omega^{\rm c}{-}\Lambda_z&-\mu\\
               -\mu&0&\mu&\Lambda_0{+}\omega^{\rm c}{+}\Lambda_z
   \end{pmatrix}
   \begin{pmatrix}
               S_1 \\
               S_2 \\
               S_3 \\
               S_4
   \end{pmatrix},
\end{equation}
where $\omega^{\rm c}=\omega\cos2\theta$
and $\omega^{\rm s}=\omega\sin2\theta$. Notice that even though we
have specialized to the homogeneous solution, we keep a possible
matter current $\Lambda_z$ that breaks the isotropy of the background
medium even though the neutrinos themselves are taken to be isotropic.

As mentioned earlier, this flavor pendulum case is easier to understand
by combining the neutrino and antineutrino modes each in a symmetric
and antisymmetric solution through $S_{1\pm4}=(S_1\pm S_4)/2$ and
$S_{2\pm3}=(S_2\pm S_3)/2$ so that
\begin{equation}\label{eq:EOM41}
  i\partial_t\begin{pmatrix}
               S_{1+4}\\
               S_{2+3}\\
               S_{1-4}\\
               S_{2-3}
             \end{pmatrix}=
   \begin{pmatrix}
               -\omega^{\rm s} \\
               \omega^{\rm s} \\
               0 \\
               0
   \end{pmatrix}+
    \begin{pmatrix}
               \Lambda_0{-}\mu{+}\omega^{\rm c}&\mu&-\Lambda_z&0\\
               -\mu&\Lambda_0{+}\mu{-}\omega^{\rm c}&0&\Lambda_z\\
               -\Lambda_z&0&\Lambda_0{+}\mu{+}\omega^{\rm c}&\mu\\
               0&\Lambda_z&-\mu&\Lambda_0{-}\mu{-}\omega^{\rm c}
   \end{pmatrix}
   \begin{pmatrix}
               S_{1+4}\\
               S_{2+3}\\
               S_{1-4}\\
               S_{2-3}
   \end{pmatrix}.
\end{equation}
If there is no matter current ($\Lambda_z=0$) the symmetric and antisymmetric
solutions decouple from each other, each forming a closed set of equations.

\paragraph{Symmetric Solution}
Assuming $\Lambda_z=0$ and with our initial conditions $S_j(0)=0$, the
antisymmetric solution will stay in this fixed point because it is not
disturbed by the mass term which has even parity.  The symmetric
solution, on the other hand, gets
triggered. If there is no
neutrino-neutrino interaction ($\mu=0$), the explicit solutions are
\begin{equation}\label{eq:pendulum-free-solution-1}
  S_{1,2}(t)=\frac{\ws}{\wc\pm\Lambda_0}
  \Bigl(1-e^{-i(\Lambda_0\pm\wc)t}\Bigr)\,,
\end{equation}
where $S_4=S_1=S_{1+4}$ and $S_3=S_2=S_{2+3}$. These are the usual
neutrino oscillations in matter, here expressed in the limit of
small off-diagonal $\varrho$ elements. The trajectory of these
solutions in the complex plane are circles of radius
$|\ws/(\wc\pm\Lambda_0)|$ and centers displaced from the origin by
the same amount, i.e., they are circles passing through the origin.

When discussing the dispersion relation of the unperturbed system we
performed a shift $K^\mu\to k^\mu=K^\mu-\Lambda^\mu$ which we can here
mimic by going to a rotating frame. In this spirit we consider the
rotating solutions $\tilde S_{1,2}(t)=S_{1,2}(t)\,e^{i\Lambda_0 t}$
that are explicitly
\begin{equation}\label{eq:pendulum-free-solution-2}
  \tilde S_{1,2}(t)=\frac{\ws}{\wc\pm\Lambda_0}
  \Bigl(e^{i\Lambda_0t}-e^{\mp i\wc t}\Bigr)\,.
\end{equation}
If we assume the limit of a large matter effect, $|\Lambda_0|\gg
|\wc|$, we may average over the fast motion and obtain
\begin{equation}\label{eq:pendulum-free-solution-3}
  \langle\tilde S_{1,2}\rangle(t)=\mp\frac{\ws}{\Lambda_0}
  e^{\mp i\wc t}\,,
\end{equation}
where $\langle\ldots\rangle$ means quantities averaged over a cycle of
the fast oscillation. These solutions correspond to a rotation in the
complex plane around the origin (which is the weak-interaction
direction) with the projected vacuum oscillation frequency
$\wc=(\Delta m^2/2|E|)\,\cos2\theta$ and initial conditions
$\langle\tilde S_{1,2}\rangle(0)=\mp\ws/\Lambda_0$.

Next we include neutrino-neutrino interactions with $\mu>0$
With the notation $\kappa=\sqrt{(2\mu-\wc)\wc}$ the full
solution is
\begin{equation}\label{eq:pendulum-solution-3}
  S_{1}(t)=\frac{\ws}{\wc}
  \[\frac{-\kappa+i\wc}{2(\kappa-i\Lambda_0)}\,e^{\kappa t}
  -\frac{\kappa+i\wc}{2(\kappa+i\Lambda_0)}\,e^{-\kappa t}
  +\frac{\kappa^2+\Lambda_0\wc}{\kappa^2+\Lambda_0^2}\,e^{i\Lambda_0t}
  \]e^{-i\Lambda_0 t}.
\end{equation}
If we assume inverted mass ordering ($\wc>0$) and $\mu>\wc/2$, the
quantity $\kappa$ is real and positive, so the first term is
an exponentially growing solution. This corresponds to the traditional
flavor pendulum solution~\cite{Hannestad:2006nj}.

One peculiar case is the limit $\wc\to0$ that obtains for maximal
mixing when $\cos2\theta\to0$. In this case $\kappa\to 0$, i.e., there
is no unstable solution. We may expand the exponentials $e^{\pm\kappa
  t}$ to lowest order in $\kappa$. For $\kappa\to 0$ we thus find the
limiting solutions $S_{1,2}(t)=(1-e^{-i\Lambda_0 t})\ws/\Lambda_0$,
corresponding to free in-medium oscillations as if there were no
neutrino-neutrino interaction. The limit of maximal two-flavor mixing
implies that the diagonal elements of the $\sM^2$ matrix are equal so
that indeed the stability analysis reveals the absence of
instabilities.

Returning to the case $\wc>0$ we now assume
a hierarchy of scales $\Lambda_0>\mu\gg\wc$ so that
$\Lambda_0\gg\kappa\gg\wc$. The asymptotic solution for
sufficiently long times for the growing modes is
\begin{equation}\label{eq:pendulum-solution-4}
  S_{1+4}(t)\simeq S_{2+3}(t)\simeq
  -i\,\frac{\ws\kappa}{2\wc\Lambda_0}
  \,e^{\kappa t}\,e^{-i\Lambda_0t}\,.
\end{equation}
One finds the same asymptotic solutions if instead of solving the equations~\eqref{eq:EOM40} with vanishing initial
conditions, one solves the corresponding equations without the constant term on the rhs but with nonzero initial coherences
$S_1(0)=S_4(0)=-\ws/\Lambda_0$ and $S_2(0)=S_3(0)=\ws/\Lambda_0$.
These initial conditions correspond to equation~\eqref{eq:pendulum-free-solution-3} at $t=0$, i.e.,
effectively we are solving the unstable solutions for quantitities
that are time-averaged over the fastest scale $\Lambda_0$.

\paragraph{Antisymmetric Solution}
The neutrino mass term cannot trigger the antisymmetric
flavor pendulum, corresponding to
$S_{1-4}$ and $S_{2-3}$. However, if the background medium
has a non-vanishing current $\Lambda_z$, we see from
equation~\eqref{eq:EOM41} that the odd and even solutions
are coupled. Therefore, after the mass has triggered the
even solution, this motion is communicated to the odd one.

To see this more explicitly we now assume normal mass ordering
($\omega<0$) where the even solution is stable, whereas the
odd one grows exponentially. We assume that $\Lambda_z$
is small so that at first the even solution develops without
backreaction from the odd one. In the limit $\Lambda_0>\mu\gg|\wc|$
we find the approximate solution
\begin{equation}\label{eq:even-oscillator}
  S_{1+4}(t)\simeq S_{2+3}(t)\simeq
  i\,\frac{\gamma \ws}{\Lambda_0\wc}\,\sin(\gamma t)\,e^{-i\Lambda_0 t}\,,
\end{equation}
where $\gamma=\sqrt{-\wc(2\mu-\wc)}$ is real for $\wc<0$. The even
solution is essentially a flavor pendulum with a small excursion and
as such a harmonic oscillator with frequency $\gamma$.

We can now solve the EOM for the odd solution, the lower right block
diagonal of equation~\eqref{eq:EOM41}, using the even solution in
equation~\eqref{eq:even-oscillator} as a driving force. With
a hierarchy of scales $\Lambda_0>\mu\gg|\wc|$, the asymptotic
solution after enough exponential growth is
\begin{equation}\label{eq:odd-oscillator}
  S_{1-4}(t)\simeq -S_{2-3}(t)\simeq
  -i\,\frac{\ws\Lambda_z}{4\wc\Lambda_0}\,e^{\kappa t}\,e^{-i\Lambda_0 t}\,,
\end{equation}
with $\kappa=\sqrt{-\wc(2\mu+\wc)}\simeq\sqrt{-2\wc\mu}\simeq\gamma$.

\subsubsection{Colliding Beams Model}

As another example of mode-triggering, we consider the homogeneous modes
with $\partial_z S_j=0$ in the general colliding beams model. In this case we
now have fast modes and
the dynamical eigenvectors of equation~\eqref{eq:fast-dyn-eigenvectors}
suggest that we should combine the neutrinos and antineutrinos of each beam
to a common mode, i.e., to consider the coherence functions
$S_{1\pm3}=(S_1\pm S_3)/2$ and $S_{2\pm4}=(S_2\pm S_4)/2$. With
this transformation the EOM for homogeneous modes is explicitly
\begin{equation}\label{eq:EOM51}
  i\partial_t\begin{pmatrix}
               S_{1+3}\\
               S_{2+4}\\
               S_{1-3}\\
               S_{2-4}
             \end{pmatrix}=
   \begin{pmatrix}
               0 \\
               0 \\
                -\omega^{\rm s} \\
               +\omega^{\rm s}
   \end{pmatrix}+
    \begin{pmatrix}
               \Lambda_0&(g_2-g_4)\mu&\wc&(g_2+g_4)\mu\\
               (g_3-g_1)\mu&\Lambda_0&-(g_1+g_3)\mu&-\wc\\
               \wc&0&\Lambda_0&0\\
               0&-\wc&0&\Lambda_0
     \end{pmatrix}
   \begin{pmatrix}
               S_{1+3}\\
               S_{2+4}\\
               S_{1-3}\\
               S_{2-4}
   \end{pmatrix}.
\end{equation}
The upper-left block-diagonal part of this EOM represents the fast modes, which
however do not have a source term from the mass matrix. However, they are
coupled to the odd modes which are excited by the mass term. We can mimic the
fast-mode-only case by using $\wc=0$ while $\ws\not=0$, a case corresponding
to maximal mixing. In this case the odd modes are not dynamical and they
are not influenced by the even ones, so their simple solution is
\begin{equation}
  S_{1-3}(t)=-S_{2-4}(t)=\frac{\ws}{\Lambda_0}\,\(1-e^{-i\Lambda_0 t}\)\,.
\end{equation}
These odd modes then seed the even modes which do not couple to the mass
term directly. For the parameters chosen in section~\ref{sec:fast-modes} (i.e., $g_1=g_2=3/2$ and $g_3=g_4=1/2$) we find
\begin{equation}
  S_{1+3}(t)\simeq -\frac{(1+i)\,\ws}{\Lambda_0+i\mu}\,e^{-i \Lambda_0 t}\,e^{\mu t}
  \quad\hbox{and}\quad
  S_{2+4}(t)\simeq \frac{(1-i)\,\ws}{\Lambda_0+i\mu}\,e^{-i \Lambda_0 t}\,e^{\mu t}
\end{equation}
for the large-$t$ asymptotic behavior.

In this example, the unstable modes could not be excited directly by the mass term, but
the disturbance was communicated through the inert modes. The eigenmodes
of the neutrino system were not orthogonal, so the excitation by the mass term
of one set of modes was communicated to those without direct overlap with the mass term. In this case a matter current was not needed to trigger these modes.

\subsection{Matter Inhomogeneity}
The main take-away message from the previous subsection is that asymmetries in the background medium are inherited by the flavor evolution. The only term responsible to kick off the evolution is the flavor-violating part of the vacuum Hamiltonian, but it can only trigger homogeneous modes by itself. Inhomogeneous modes get triggered only when we break homogeneity of the background medium by introducing some non-uniformity in the matter density.

To see this effect explicitly we need to consider the time evolution of inhomogeneous modes separately, and for that we need to look at the Fourier transform of the EOMs over space in
the presence of time-independent small density fluctuations $\Lambda_0+\delta\Lambda_0(\br)$,
where $\Lambda_0\gg \delta\Lambda_0(\br)$ is the homogeneous matter density
and for simplicity we assume that the background
current $\mathbf{\Lambda}=0$.
We denote the amplitude of each $\textbf{k}$ mode at time $t$ by $S_{E,\textbf{v}}(\textbf{k},t)$, so the transformed EOMs are
\begin{multline}
i \partial _t S_{E,\textbf{v}}(\bk,t) - (\textbf{v} \cdot \textbf{k})S_{E,\textbf{v}}(\bk,t) = -\omega_E^{\rm s}\delta(\textbf{k}) + \Lambda_0  S_{E,\textbf{v}}(\textbf{k},t)\\
+ \int d \textbf{k}' \,
\delta\Lambda_0 (\textbf{k} - \textbf{k}') S_{E,\textbf{v}}(\textbf{k}', t) + \omega_E^{\rm c} S_{E,\textbf{v}}(\textbf{k},t)
- v^\alpha \int d \Gamma' v'_\alpha g_{E',\textbf{v}'} S_{E',\textbf{v}'}(\textbf{k}, t)  ,
\end{multline}
where $\delta\Lambda_0(\bk)$ are the Fourier modes of the small density
variation.

It is clear from this equation that matter variations couple different $\textbf{k}$ modes which makes it possible to trigger non-zero $\textbf{k}$ modes. We now drop the self interaction term which is a good approximation for small $t$. We also omit the subscript $\{E,\textbf{v}\}$ from now onwards.  Now we can find the solution to this initial-value problem for small times perturbatively by writing the transformed EOMs as
\begin{multline}\label{eq:inhomogeneous-trigger}
e^{i\textbf{v} \cdot \textbf{k}t} S(\textbf{k},t) =  i\omega_E^{\rm s} \delta(\textbf{k}) t \\
- i \left[ \int_0^t dt' e^{i\textbf{k} \cdot \textbf{v} t'}(\Lambda_0+\omega^{\rm c}) S(\textbf{k}, t') +  \int_0^t dt' \, e^{i \textbf{k} \cdot \textbf{v} t'} \int d \textbf{k}' \,
\delta\Lambda_0(\textbf{k} - \textbf{k}') S(\textbf{k}', t') \right].
\end{multline}
Initially the $S$-functions are zero, hence only the homogeneous term is excited which then couples to the rest of the modes and excites them too. Given that the coupling between different $\textbf{k}$ modes is small we can safely assume that there is no feedback. So, to the zeroth order
\begin{equation}
S(\textbf{k}, t) = \frac{\omega^{\rm s}(1 - e^{-i(\Lambda_0 + \omega^{\rm c})  t})}{\Lambda_0 + \omega^{\rm_c}} \delta(\textbf{k})\,.
\end{equation}
Putting this back into equation~\eqref{eq:inhomogeneous-trigger} for $\textbf{k} \neq \textbf{0}$ we find
\begin{equation}
S(\textbf{k}, t) = \frac{\omega^{\rm s}}{\Lambda_0 + \omega^{\rm c}} \delta \Lambda_0 (\textbf{k}) \int_0^t dt'\,e^{i\textbf{v} \cdot \textbf{k}(t'-t)}(1 - e^{-i(\Lambda_0 + \omega^{\rm c}) t'})\,,
\end{equation}
which shows how the variations in background matter density couple the instability in a homogeneous mode to the inhomogeneous modes.

\subsection{Section Summary}

These explicit examples show that the logic behind the linearized
stability analysis remains justified for non-vanishing neutrino mixing,
implying non-vanishing off-diagonal elements of the mass
matrix. However, we need a large matter effect which implies that
without neutrino-neutrino interactions, all polarization vectors
remain in a narrow cone around the flavor direction. The fixed-point
solution is actually fuzzy and not a fixed point, but a small
environment around $S_j=0$. It is this fuzziness of the ``fixed
point,'' caused by the off-diagonals of $\sM^2$, which triggers the
unstable solutions.

However, the mass term is perfectly symmetric
(static, translational invariant, isotropic, parity even, etc.) and
as such can only affect collective modes with the same symmetries.
On the other hand, it is enough that the background medium through
the matter effect violates these symmetries to couple the mass term
to other unstable modes. It is ultimately the interplay of the mass
term with the background medium that triggers collective modes of
neutrino flavor coherence.

\section{Conclusion}
\label{sec:conclusion}

Starting from the usual kinetic equation for the flavor content of a
locally homogeneous neutrino gas, we have studied the EOM for the
two-flavor coherence functions in the linear regime. Under the
assumption of vanishing neutrino mixing, there is a fixed point
corresponding to vanishing flavor coherence, serving as the starting
point for the linearization.  Based on the flavor-dependent neutrino
energy and angle distributions we have derived the dispersion relation
for collective modes that emerge in the presence of neutrino-neutrino
interactions measured by $\mu=\sqrt{2}\GF n_\nu$.  Including neutrino
mass differences (but still ignoring flavor mixing) the dispersion
relation can show fast modes, which are dynamical even for $\Delta
m^2=0$ and thus $\omega=\Delta m^2/(2E)=0$, and/or slow modes which
require $\omega\not=0$. Either type of mode typically shows unstable
behavior for some range of frequencies (imaginary wavevector) and/or
some range of wavevectors (imaginary frequency). The imaginary part
can be of the order of $\mu$ (fast instability) or $\sqrt{|\omega
  \mu|}$ (slow instability). Slow modes can only have slow
instabilities, whereas fast modes can have fast or slow instabilities.

Up to this point we have studied a wave equation without
sources. Within standard physics, the only source of flavor coherence
is flavor mixing by the neutrino mass term. It destroys the fixed
point of the original EOM, so we need to include a large matter effect
which essentially de-mixes neutrinos in that propagation eigenstates
nearly coincide with flavor eigenstates. Therefore, without
neutrino-neutrino interactions the solutions are flavor oscillations
with a very small amplitude. Therefore, our linearization of the EOM
remains strictly correct, whereas the original fixed point becomes
fuzzy in that modes with different $E$ precess with different
frequencies and around different center points in the complex plane of
flavor coherence. Unstable modes are thus seeded by the mass term
and, after sufficient growth, coincide with the unstable modes derived
from the dispersion relation.

We have studied explicit examples and general expressions for the
triggering of unstable modes. If they have
non-vanishing wavenumber they require fluctuations of the
matter term to mediate between the homogeneous and isotropic mass term
and unstable modes that break these symmetries. We have derived an
explicit expression for the seed amplitude of a given unstable mode
based on the given amplitude of matter fluctuations. In this context
we have stressed that the spectrum of collective modes is not
complete---there can be modes that remain non-dynamical (inert) even in the
presence of neutrino-neutrino interactions, although our dispersion
relation captures all possible unstable modes. Moreover, the
propagating and/or unstable collective modes, as well as the inert ones, need not
be orthogonal. The inert modes can be crucial for mediating the
disturbance caused by the mass term to the exponentially growing modes.

The key limitation of our analysis is that it only predicts if and 
when an instability can occur, but not what the final outcome will be. 
Specifically, this depends on what seeds the instability and this 
information is perhaps system-dependent. Also, the analysis is limited to 
a local volume inside which conditions are taken to be static and homogeneous 
(except due to the effect of instabilities under consideration) but extending our analysis to 
the situation where the external conditions are varying requires new ideas. Other 
ingredients, e.g., including collisions, external forces such as gravity, and 
effects of time-of-flight, can be challenging 
to implement within this framework. On the other hand, extending the analysis presented 
in this paper to include the spin, spin-flavor, and neutrino-antineutrino correlations, 
is perhaps more straightforward.

\section*{Acknowledgments}

In Munich, we acknowledge partial support by the Deutsche Forschungsgemeinschaft through Grants No.\ EXC 153 (Excellence Cluster ``Universe'') and
SFB 1258 (Collaborative Research Center ``Neutrinos, Dark Matter,
Messengers''), as well as the European Union through Grant
No.\ H2020-MSCA-ITN-2015/674896 (Innovative Training Network
``Elusives''). S.A.\ acknowledges support by a WISE Fellowship of the
Deutscher Akademischer Austauschdienst (DAAD) for a
summer internship at MPP. The work of B.D.\ is partially supported by the
Dept.\ of Science and Technology of the Govt.\ of India through a
Ramanujan Fellowship and by the Max-Planck-Gesellschaft through
a Max-Planck-Partnergroup. S.C. acknowledges the  support of the
Max Planck India Mobility Grant from the Max Planck Society, 
supporting the visit and stay at MPP.

\end{document}